\documentclass[twocolumn]{revtex4}

\def\be{\begin{equation}}
\def\ee{\end{equation}}
\def\la{\label}
\def\bea{\begin{eqnarray}}
\def\eea{\end{eqnarray}}
\def\non{\nonumber}
\def\ci{\cite}
\def\la{\label}
\def\bib{\bibitem}

\def\le{\left}
\def\ri{\right}

\def\Omp{\Omega_\phi}

\def\rp{\rho_\phi}
\def\rpo{\rho_{\phi o}}
\def\wp{w_\phi}

\def\Ompo{\Omega_{\phi o}}

\def\s8{\sigma_8}

\def\fr{\frac}

\def\non{\nonumber}

\def\Omp{\Omega_\phi}

\def\r{\rho}

\def\rp{\rho_\phi}
\def\rpi{\rho_{\phi i}}
\def\rpo{\rho_{\phi o}}

\def\rb{\rho_b}

\def\rbo{\rho_{b o}}

\def\wp{w_\phi}

\def\Ompo{\Omega_{\phi o}}

\def\we{w_{eff}}
\def\wbe{w_{b eff}}
\def\wa{w_{app}}

\usepackage{graphicx}

\begin{document}

\title{The Fate of the Universe: Dark Energy Dilution? }

\author{A. de la Macorra }
\affiliation{Instituto de F\'{\i}sica UNAM\\ Apdo. Postal 20-364\\
01000 M\'exico D.F., M\'exico}

\begin{abstract}

We study the possibility that dark energy  decays in the   future
and  the universe stops accelerating. The fact that
the cosmological observations  prefer an equation of state of dark energy
smaller than -1
can be a signal that dark energy will
decay in the future. This conclusion is based in interpreting
a $w<-1$  as a signal of dark energy interaction
with another fluid. We determine the interaction through
the cosmological data  and extrapolate  it into the future.
The resulting energy density for dark energy becomes
$\rp=a^{-3(1+\wp)}e^{-\beta(a-1)}$, i.e.  it has
an exponential suppression for $a\gg a_o=1$. In this scenario
the universe ends up dominated by this other fluid, which could be
matter, and the universe stops  accelerating at some time in the
near future.

\end{abstract}


\maketitle

\section{Introduction}

In the last few years the existence of dark energy as a fluid with
negative pressure that accelerates the universe at present time has
been established \ci{DE},\ci{SN}. Within the context of field theory
and particle physics
 it is appealing  to interpret  the dark energy as some kind
 of particles that interact with the particles of the standard model
 very weakly. The weakness of the interaction is required since dark
 energy particles have not been produced in the accelerator
and because
 the dark energy has not decayed into lighter (e.g. massless) fields
such as the photon. Perhaps the most appealing candidate for   dark
energy is that of a scalar field, quintessence \ci{Q}, which can be
either a fundamental particle or a composite particle \ci{Qax}. It
was common to assume the interaction between the dark energy and all
other particles to be via gravity only, however recently interacting
dark energy models have been proposed \ci{IDE}-\ci{wapp}. The
interesting effect of this interaction is two fold. On the one hand,
the interaction between dark energy and matter, which can be for
example dark matter  or neutrinos \ci{IDE-n}, is to give an apparent
equation of state of dark energy   smaller (more negative) than
without the interaction and can be even smaller than -1 \ci{wapp},
as suggested by the cosmological observations. On the other hand it
is also possible to have dark energy interacting with neutrinos and
it is tempting to relate both energies since they are of the same
order of magnitude \ci{IDE-n} and a mass of neutrinos larger than
$0.8 eV$ imply that the dark energy cannot be a cosmological
constant \ci{IDE-ax}.

In general fluids with $w<-1$ give
many theoretically problems such as stability issues or wrong
kinetic terms as   phantom fields \ci{ph.etc}. However,  interacting dark energy is a very
simple and attractive  option which we will use in this
{\it letter}.

Since the dark energy dilutes slower than matter we expect it to
dominate the universe at late times. So, once the universe begins to
accelerate due to dark energy we expect it to maintain this state of
acceleration in the future and the universe will end  up completely
dominated by dark energy. In this {\it letter} we would like to
study if this fate of the universe is unavoidable or we could have a
transition from an accelerating universe to a non accelerating one
in which the dark energy decays  into another fluid. We will show
that the fact that the cosmological data \ci{DE}, specially the SN1a
data \ci{SN}, prefer an equation of state $w$ of dark energy smaller
than minus one can be a signal that dark energy will decay in the
future and the universe will stop accelerating. This conclusion is
based in interpreting a $w<-1$  as a signal of dark energy
interaction with another fluid. We determine  the interaction
through the cosmological observations  and extrapolating it into the
future.

This {\it letter} is organized as follows. In section \ref{ide}
we present the generic evolution of two interacting fluids.
In section \ref{eff} we introduce the effective and apparent
equations of state and in section \ref{ded} we give
  the evidence  a dark energy decay. Finally we present
  our conclusions.

\section{Interacting Dark Energy}\la{ide}

\subsection{Fluid Evolution}

The evolution of two interacting fluids $\rp$ and $\rb$, which can be quintessence scalar field
$\phi$ for dark energy "DE" and another fluid ($\rb$), as for example matter or
radiation, is given by
\bea\la{rpp}
\dot\rp&=& - 3H\rp(1+\wp)-\delta(t)\\
\dot\r_b&=&  - 3H\rb(1+w_b)+\delta(t)
\la{rpb}\eea
with $H=\dot a/a$ the Hubble parameter and $\delta (t)$ the  interaction coupling.
This $\delta$  is a dissipative term,
it depends on the interaction term    between the particles
of $\rp$ and $\rb$ and is in general a function of time.
The equation of state parameters
$w_\phi\equiv p_\phi/\rp $ and $w_b\equiv p_b/\rb $ may
 be functions of time. Without lack of generality we will take $w_b>\wp$,
 which is consistent with assuming $\rp$ as the dark energy
 in the absence of an interaction term.

A simple general solution for $\rb$ can be obtained by taking  $w_b$
constant and  $\delta = A(t) \rb$.  Eq.(\ref{rpb}) becomes $\dot\rb=
-[3H(1+\wp)-A]\rb$ and it has a solution
\be\la{srb}
\rb = \rbo a^{-3(1+w_b)}\,e^{\int_{to}^t dt A}=  \widetilde{\rb}\,e^{\int_{to}^t  dt   A}
\ee
with
$\widetilde{\rb}\equiv \rbo a^{-3(1+w_b)}$, the evolution of $\rho_b$
without an interacting term, i.e. $\delta =0$. We take $a(t_o)\equiv a_o=1$ at
present time. A similar solution can be obtained for $\rp$ with $\wp  $
constant and setting $\delta = \widetilde{A} \rp$
the solution to eq.(\ref{rpp}) gives
\be\la{srp}
\rp = \rpo a^{-3(1+\wp)}\,e^{-\int_{to}^t  dt  \widetilde{A}}=
\widetilde{\rho}_\phi \,e^{-\int_{to}^t  dt  \widetilde{A}}
 \ee
 with  $\widetilde{\rho}_\phi\equiv \rpo a^{-3(1+\wp)}$
the evolution  for $\delta =0$.
Of
course $A,  \widetilde{A}$ are related by  $\delta=
 \widetilde{A} \rp =  A \rb$. Clearly the sign of
 $A$, i.e. $\delta$, determines whether $\rp$ and $\rb$
 will  evolve faster or slower with  the interaction term
 than without it.  So, for $A$ positive $\rb$ will dilute slower while
 for $A$ negative it will dilute faster.

If we take the ratio $y\equiv \rp/\rb=\Omp/\Omega_b$ from eqs.(\ref{srb})
and (\ref{srp}) we have
\be\la{y}
y=\fr{\rp}{\rb}=y_o a^{3(w_b-\wp)}\,e^{-\int_{to}^t  dt  A(1+1/y)}
\ee
with $y_o=\rpo/\rbo$,
and taking the derivative of $y$ w.r.t. time we find
\be\la{yt}
\dot y= 3Hy\le[\Delta w  - \Upsilon\ri]
\ee
with $\Delta w \equiv w_b-\wp$ and
\be\la{u}
\Upsilon\equiv \fr{\delta}{3H}\le(\fr{\rp +\rb}{\rp\rb}\ri)=\fr{A (1+y)}{3Hy}.
\ee
The value for $y$ is constraint to
$0\leq y\leq \infty$ with $y=0$ for $\rp=0$ and
$y=\infty$ for $\rb=0$.
Clearly from eq.(\ref{yt}) we see that
the evolution of $y$ depends on the sign of $\Delta w - \Upsilon$.

\subsubsection{Non Interaction solution: $\Delta w > \Upsilon $}\la{ni}

If $(w_b-\wp)> \Upsilon $ then $\dot y$ is positive and
$y$ will increase, i.e. $\rb$ will dilute  faster than $\rp$,
and we will end up with $\rp$ dominating the universe. In this
case the interaction term $\delta$ is subdominant and the evolution
of $\rp$ and $\rb$ is the usual one, i.e.
$\rp \propto a^{-3(1+\wp)}$ and $\rb\propto a^{-3(1+w_b)}$.

\subsubsection{Interacting solution: $\Delta w < \Upsilon $}\la{in}

For $\rb$ to dominate the universe we need  $y\ll 1$ at late time
and the (interaction) term $\Upsilon$ should dominate over $\Delta w$.
A simple example is when $\widetilde{A}$ is constant and positive. In this case
eq.(\ref{y})  is
\be
y=y_i\le(\fr{a}{a_i}\ri)^{3(w_b-\wp)}\,e^{- \widetilde{A} t} \rightarrow 0
\ee
at late times for any value of $w_b-\wp$.

\subsubsection{Finite solution: $\Delta w=\Upsilon $}\la{=}

A  solution to eq.(\ref{yt}) with $y$ constant and
$\rb\neq 0, \rp\neq 0$ is only possible if $A/H$ (or $\widetilde{A}/H$) is positive and constant since
$\Upsilon= A (1+y)/3Hy= \widetilde{A} (1+y)/3H=\Delta w$ must be constant and positive,
taking $w_b-\wp>0$ constant. Let us take
$\widetilde{A}/H=C>0$  constant, i.e. with $\delta=\widetilde{A}\rp= C H \rp$,
and from $\dot y=0$, i.e. $\Upsilon= \widetilde{A} (1+y)/3H=\Delta w$,
we get a stable value of $y$ given by
\be
y_s =\fr{\Omp}{\Omega_b}= \fr{H}{\widetilde{A}}\;3(w_b-\wp)-1=\fr{3(w_b-\wp)}{C}-1.
\ee
It is easy to see that the solution $y_s$ is stable
since from eq.(\ref{yt}) the fluctuation $\delta y= y-y_s$ to first order
behaves as $\delta \dot y/\delta y =-\widetilde{A} y_s $ ($\widetilde{A}$ is
positive by hypothesis) giving $\delta y\rightarrow 0$. Furthermore, since $\widetilde{A}$
is proportional to $H=\dot a/a$
then  we can integrate eq.(\ref{srp}) with
$\widetilde{A} dt=[3(w_b-\wp)/(1+y_s)]H dt=C H dt = C da/a$ to give
\be\la{rp3}
\rp =\rpi \le(\fr{a}{a_i}\ri)^{-3(1+\wp)-C}=
\rpi \le(\fr{a}{a_i}\ri)^{- 3(1+\fr{w_b+y_s\wp}{1+y_s})}
\ee
where we have used that $C=3(w_b-\wp)/(1+y_s)$.
Since $C$ is positive the solution in eq.(\ref{rp3})
dilutes faster than the non interacting solution
$\rp \propto a^{-3(1+\wp)}$ and with an effective equation of state
$w_{eff}=\wp+C/3=(w_b+y_s\wp)/(1+y_s)>\wp$.

\section{Effective and Apparent Equation of State}\la{eff}

We will now introduce the effective eq. of state $\we$
and the apparent eq. of state $\wa$.

\subsection{Effective  Equation of State}

To obtain an effective equation of state we simply
rewrite eqs.(\ref{rpp}) and (\ref{rpb}) as
\bea\la{rw}
\dot\rp &=& -3H\rp(1+w_{eff} )\non\\
\dot\rb &=& -3H\rb(1+w_{b eff})
\eea
with the effective equation of state defined by
\be\la{weff}
w_{ eff}  = \wp+ \fr{\delta}{3H\rp},\hspace{1cm}
w_{b eff} = w_b-\fr{\delta}{3H\rb}.
\ee
The solution to eq.(\ref{rw}) is
\bea\la{rw2}
\rp &=& e^{-3\int_{to}^t  (1+w_{eff} )da/a}=a^{-3(1+\wp)}\,e^{-\int_{to}^t \fr{\delta}{\rp} dt}\non\\
\rb &=& e^{-3\int_{to}^t  (1+w_{b eff} )da/a}=a^{-3(1+w_b)}\,e^{\int_{to}^t \fr{\delta}{\rb} dt}
\eea
where we have assumed in the second equality of eqs.(\ref{rw2}) that $\wp,w_b$ are constant.
We see from eqs.(\ref{rw2}) that $\we,\wbe$ give the complete evolution of
$\rp$ and $\rb$.
For $\delta>0$ we have $w_{ eff}>\wp$ and
the fluid $\rp$ will dilute faster then without the interaction
term  (i.e. $\delta=0$) while $\rb$ will dilute slower since $w_{b eff}<w_b$.
Which fluid dominates at late time will depend on which
effective equation of state is smaller. The difference in
eqs.(\ref{weff}) is
\be\la{dw}
\Delta w_{eff}\equiv w_{b eff}-w_{ eff}=\Delta w - \Upsilon
\ee
with $\Upsilon$ defined in eq.(\ref{u})
while the sum gives
\be\la{sw}
\Omega_bw_{b eff} +\Omp\we= \Omega_bw_b+\Omp\wp.
\ee
Clearly the relevant quantity to determine the relative growth
is given by $\Upsilon$
and if   $\Upsilon > \Delta w$
 we have  $\Delta w_{eff}<0$ and $\rb$ will dominate the universe at late times
 while for $\Upsilon < \Delta w$ we have  $\Delta w_{eff}>0$ and
 $\rp$ will prevail. For no interaction $\delta=0$  and  $\Upsilon=0$ giving
$\Delta \we=\Delta w>0$ and $\rp$ dominates at late times.
If $\Upsilon = \Delta w$ then $w_{b eff}=w_{ eff}$ and
 the ratio of both fluids $\rho_b/\rp$ will approach a constant value, and
if the universe is dominated by
$\rp+\rb$, i.e. $\Omp+\Omega_b=1$, then eq.(\ref{sw}) gives
\be
\wp\leq\; w_{eff}= w_b\Omega_b+\wp(1-\Omega_b) \;\leq w_b,
\ee
i.e. the effective equation of state is constraint between $\wp$ and $w_b$.
Of course eqs.(\ref{dw})
  and (\ref{u}) are consistent with the analysis of
eq.(\ref{yt}).

\subsection{ Apparent Equation of State}

 An interesting result of the interaction between   dark
 energy   with other particles
 is to change  the apparent
 equation of state of dark energy
 \ci{IDE}-\ci{wapp}. An observer that supposes
 that DE has no interaction sees a different
 evolution of DE as an observer that takes into account
for the interaction between DE and another fluid. This effect allows
to have an  apparent equation of state $w<-1$ for the
``non-interaction" DE \ci{wapp}  even though the true equation of
state of  DE is larger than -1.

Let as take the
 energy density   $\rho=\rp+\rb=\rho_{DE}+\widetilde{\rb}$.
 The energy densities $\rp,\rb$ are given by eqs.(\ref{rpp})
 and (\ref{rpb}) and these two fluid interact via  the
 $\delta $ term. On the other hand the energy
 densities $\rho_{DE}$ and $\widetilde{\rb}$ do not interact with each other
 by hypothesis and  therefore we have
$\dot{\widetilde{\rb}}=-3H(1+w_b)$ and
$\dot\rho_{DE}=-3H\rho_{DE}(1+w_{ap})$, i.e.
\bea
\widetilde{\rb}&=&\rbo a^{-3(1+w_b)}\non\\
\rho_{DE}&=&\rho_{DE o} a^{-3(1+w_{ap})}
\eea
 if $w_b, \wa$ are constant.
It was pointed out  that the apparent equation of state $w_{app}$
can take values smaller than -1 and it is given by \ci{wapp}
\bea\la{wap}
  w_{ap}&=&\fr{\wp}{1-x}\\
x\equiv -\fr{\widetilde{\rb}}{\rp}\le(\fr{\rb}{\widetilde{\rb}}-1\ri)
&=& -\fr{\rbo a^{-3}}{\rp}\le(e^{\int_{to}^t  dt A}-1\ri)\non
\eea
valid if  $w_b=0$ and we have used eq.(\ref{srb}).
We see from eq.(\ref{wap})
that for $\rb<\widetilde{\rb}$, i.e.   $A>0$ (or $\delta>0$),
we have $x>0$ and $w_{ap} < \wp$ for $t<t_o$
which allows to have a $w_{ap}$ smaller than -1.

The result in eq.(\ref{wap}) can be generalized to the interaction
between two arbitrary  fluids.
Taking the time derivative of $\rho=\rp+\rb$ and using    eqs.(\ref{rpp}) and (\ref{rpb}),
for an arbitrary
interacting term $\delta$, we get
$
 \dot\rho_{DE}=-3H\rho_{DE}(1+w_{ap})
$
 with
 \be\la{wap2}
w_{ap}=\fr{\wp-w_bx}{1-x}
 \ee
with
\be\la{x}
x\equiv -\fr{\widetilde{\rb}}{\rp}\le(\fr{\rb}{\widetilde{\rb}}-1\ri)
= -\fr{\rbo a^{-3(1+w_b)}}{\rp}\le(e^{\int_{to}^t  dt A}-1\ri)
\ee
where we have used eq.(\ref{srb}).
Of course $w_{ap},\,
x$ in eq.(\ref{wap2}) reduce for $w_b=0$ to $w_{ap},\, x$  as given
in eq.(\ref{wap}).
For $t=t_o$ we have $x=0$  and if there is no interaction
$\delta=A=0$ we have $\rb=\widetilde{\rb}$, $x=0$ and $w_{ap}=\wp$.
An apparent equation of state $w_{ap} < \wp $  is given for $0< x\Delta w=x(w_b-\wp)$,
i.e. $x>0$.
A positive  $x$ needs $ \widetilde{\rb}<\rb$
which from eq.(\ref{srb}) implies a positive interaction term $\delta=A\rb$.
We also see that a positive $w_b$   gives a more
negative  $ w_{ap}$ than for $w_b=0$.

It is interesting to note, see table \ref{t1}  that the apparent equation of
state $w_{app}$ is smaller
than $\wp$ for a positive $\delta$ while
the effective equation of state $w_{eff}$ is in this case
larger than $\wp$. This clearly shows that the apparent
equation of state is an "optical" effect not a true
evolution.

\begin{table}
\begin{center}
\begin{tabular}{|c|c|c|c|}
  \hline
  $\delta (t)$ & $\we, \forall a$ &   $w_{app}(a<a_o)$ &   $w_{app}(a>a_o)$  \\  \hline
  $\delta > 0 $ &   $w_{eff}>\wp$ & $w_{app}<\wp$ & $w_{app}>\wp$  \\
  $\delta < 0$ &  $w_{eff}<\wp$  & $w_{app}>\wp$  & $w_{app}<\wp$ \\
  \hline
\end{tabular}
\end{center}
\caption{\small{We show the different relative sizes of $w_{app}$ and $\we$
with respect to $\wp$ as a function of the sign of the interaction term  $\delta (t)$. }}
\la{t1}\end{table}

\section{Evidence for Dark Energy Decay }\la{ded}

The SN1a observations prefer an equation of state $w<-1$ for dark energy.
In principal a $w<-1$ for a fluid is troublesome since it has
instabilities and causality problems. However, as seen in section
\ref{eff} this can be an optical effect due to the interaction between
dark energy with other particles as for example dark matter or neutrinos.

Here, we will assume that $w<-1$ due to this interaction and we will
show that this can be interpreted
as a signal for a dark energy decay in the future with
the universe  no longer accelerating.

The complete evolution of dark energy is given by the  effective equation
of state $\we$ given by eq.(\ref{weff})  while the observed
equation of state, if we assume no interaction, is given by eq.(\ref{wap}),
 i.e. $w_{app}=\wp/(1-x)$.

For values of the scale factor $a$ close to present day
$a_o=1$  we propose to approximate $x$ linearly  by
\be\la{xa}
x=\beta (a_o-a)=-\beta \delta a
\ee
with $\beta$ a constant to be determined by observations
and $\delta a\equiv a-a_o$.
A positive $x$ in the past (i.e. $a<a_o$) requires
$\beta > 0$. Eq.(\ref{xa}) satisfies
the requirements $x(a_o)=0$ and we have for $\wa>\wp$,  $x<0$ for $a>a_o$.

The SN1a data are in the range $1>a>2/5$, i.e. for a redshift
$0<z<1.5$, and the best fit solution has  an average equation of
state $<w>\approx -1.1$ \ci{DE}. Taking the average of
$w_{app}=\wp/(1-x)$ we have
 \be\la{wa2}
 <\wa>\equiv
\fr{\int_{a_1}^{a_o} \wa\,da}{\int_{a_1}^{a_o}da}= \fr{\wp
Log\le[1-\beta(1- a_1)\ri]}{\beta (a_1-1)}
\ee
where we have used
eq.(\ref{xa}) and $a_o=1$. As an example let as take $<\wa>=-1.1$,
as suggested by the observations \ci{SN},\ci{IDE-ax}, and $ \wp=-0.9,w_b=0, \;a_1=2/5 $.
In this case we obtained from eq.(\ref{wa2}) the value $\beta=0.56$.
If instead of taking the average as in eq.(\ref{wa2}) we simple take
the $\wa$ evaluated at the extreme points we have
$<\wa>=\fr{1}{2}(\wa(a_o)+\wa(a_1))$ giving the simple analytic
expression $\beta=\fr{2<\wa>-2\wp}{(1-a_1)(2<\wa>-\wp)}$ and for our
previous example we find $\beta=0.51$. The difference in determining
$\beta$ is small and the expression for $\beta$ becomes analytic as
a function of $<\wa>$.

Now, we would like to determine the effective (true) eq. of state
for  dark energy. From eq.(\ref{wap}) we have
\be\la{x2}
x =-\fr{\rho_b}{\rp}\le(1-e^{-\int_{to}^t dt A} \ri)
\simeq -\fr{\rho_b}{\rp} A\delta t
\ee
where we have approximated $e^{\int_{to}^t dt A}\simeq 1-A\delta t$
and $\delta t\equiv t-t_o$.
Taking $\delta=A\rb$ and eq.(\ref{x2}) we have
$\delta=-x\rp/\delta t$   and
using
$\delta t= \delta a/aH$ and eq.(\ref{xa}) we get an interaction term
\be\la{d}
\delta=a\beta H\rp.
\ee
Eq.(\ref{weff}) becomes then
\be\la{wex}
\we=\wp+ \fr{\delta}{3H\rp}\simeq
  \wp+ \fr{a \beta}{3 }
\ee
and the effective equation of state for the b-fluid gives
\be\la{wbe}
\wbe=w_b - \fr{\delta}{3H\rb}\simeq w_b- \fr{a \beta}{3 }\fr{\Omp}{\Omega_b}.
\ee
Using eq.(\ref{srp}) or equivalently eq.(\ref{rw2}) with the interaction
term given in eq.(\ref{d}) we get an energy density
\be\la{rp2}
\rp=a^{-3(1+\wp)}e^{-\beta(a-1)}
\ee
which shows that $\rp$  dilutes as $a^{-3(1+\wp)}$ for $a\ll a_o=1$
and $\rp$ is exponentially suppressed for $a \gg 1$.
From eqs.(\ref{wex}) and (\ref{wbe}) we find that
$\wbe<\we$ for $a>a_*\equiv 3 \Delta w\Omp/\beta$ (taking $\Omp+\Omega_b=1$),
i.e. for $a> a_*$ dark energy dilutes faster than the b-fluid.

For our previous example with $\wp=-0.9,w_b=0, \beta=0.56$ and  present day values
$\Ompo=0.7$ and $\Omega_{b o}=0.3$, we find
$\we=\wbe$ at $a_*\simeq 1.3$. Furthermore
$\we=0$ at $a=-3\wp/\beta=4.8 $, i.e. due to the interaction $\we$ grows from
$\we=-0.9$  to $\we=0$. In this case
the universe goes from a decelerating to an accelerating
epoch and back to a decelerating one at a scale factor $a_{d}=5.1$.

From eq.(\ref{wbe}) with $w_b=0$
we have  $w_{b eff} \leq 0$.
Since $w_{b eff}< \we$ for $\Omp\neq 0$ and $a>a_*$ then we will have at late times
$\Omp\rightarrow 0$ and $w_{b eff}\rightarrow 0$ with a universe
completely dominated by the fluid $\r_b$, matter in this case,
and decelerating.

We show in fig.\ref{fig1}  the evolution of $\Omp, \Omega_b$ in the
example with an interaction term $\delta=-x\rp/\delta t=-a\beta
H\rp$ with $x$ given by eq.(\ref{xa}) and $\wp=-0.9$ and
$\beta=0.56$, such that the average $<\wa>=-1.1$. Notice that for
$a<1$ the dark energy grows relative to the b-fluid while for $a>1$
(i.e. in the future) the b-fluid dominates. In fig.\ref{fig2} we
show the effective equations of state given by eqs.(\ref{wex}) and
(\ref{wbe}) and both are larger than -1 at all times. In
fig.\ref{fig3} we show the apparent eq. of state (c.f.
eq.(\ref{wap})) where $\wa$ is smaller than $\wp$ for $a<1$ as
suggested by the SN1a data but it becomes larger than $\wp$ for
$a>1$.

\begin{figure}[htp!]
\begin{center}
\includegraphics[width=7cm]{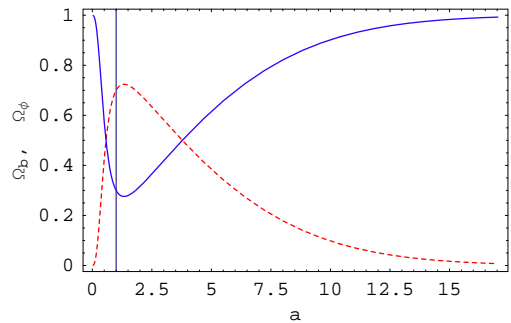}
\end{center}
\caption{\small{We show  the energy densities  $\Omp,\Omega_b$
(red (dotted) and blue (solid) respectively) as a function of the scale factor $a$.
The $b$-fluid dominates at late time. The vertical line is present time $a=1$.
}}
\la{fig1}
\end{figure}

\begin{figure}[htp!]
\begin{center}
\includegraphics[width=7cm]{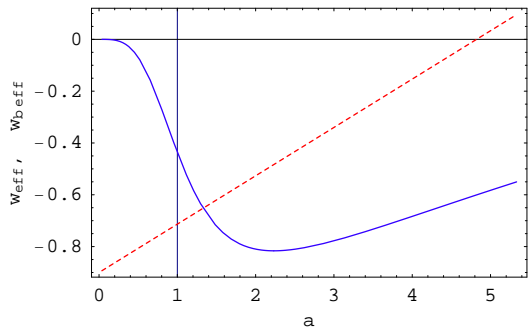}
\end{center}
\caption{\small{We show  the effective eqs. of state $\we,\wbe$
(red (dotted) and blue (solid)  respectively) as a function of the scale factor $a$.
The vertical line is present time $a=1$.
}}
\la{fig2}
\end{figure}

\begin{figure}[htp!]
\begin{center}
\includegraphics[width=7cm]{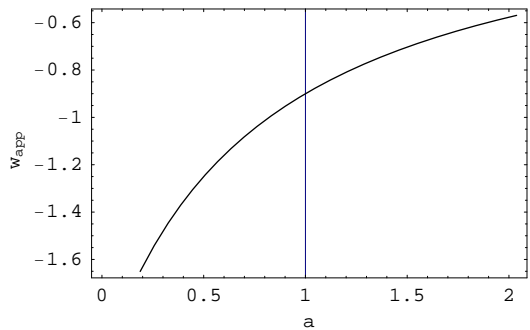}
\end{center}
\caption{\small{We show  the apparent eq. of state $\wa$
as a function of the scale factor $a$. It is less than -1
for $a<a_0=1$. The vertical line is present time $a=1$.
}}
\la{fig3}
\end{figure}

\section{Conclusions}

The cosmological observations prefer an equation of state $w<-1$ for dark energy.
We obtain an apparent equation of state for dark energy smaller
than minus one due to the interaction between dark energy and another
fluid (b-fluid). Form the observational data we determine the
interaction term, close to present day, and we show
that this interaction imply that dark energy will dilute
faster than the b-fluid. The interaction term
is $\delta=a\beta H\rp$ which gives an energy
density $\rp=a^{-3(1+\wp)}e^{-\beta(a-1)}$,   it has
an exponential suppression for $a\gg a_o$.
The resulting universe is a decelerating universe
 dominated by the b-fluid, which  could be
dark matter.

\begin{acknowledgments}

This work was also supported in
part by CONACYT project 45178-F and DGAPA, UNAM project
IN114903-3.

\end{acknowledgments}


\begin{thebibliography}{99}

\bib{DE}
D. N. Spergel {\it et al.},
 astro-ph/0603449;
M. Tegmark et al.  Phys.Rev.D74:123507,2006;
 U.~Seljak, A.~Slosar and P.~McDonald,JCAP 0610:014,2006

\bib{SN}  A.~G.~Riess {\it et al.}  [Supernova Search Team
Collaboration],
  Astrophys.\ J.\  {  607}, 665 (2004);
 W.~M.~Wood-Vasey {\it et al.},
  arXiv:astro-ph/0701041;
  N.~Palanque-Delabrouille  [SNLS Collaboration],
  arXiv:astro-ph/0509425.





\bib{Q} I. Zlatev, L. Wang and P.J. Steinhardt, Phys. Rev.
Lett.82 (1999) 896;  Phys. Rev. D59 (1999)123504; A. de la Macorra,
G. Piccinelli { Phys.Rev.D61:(2002)123503}




\bib{Qax} P. Binetruy, Phys.Rev. D60 (1999) 063502, Int. J.Theor.
Phys.39 (2000) 1859; A. de la Macorra, C. Stephan-Otto, {
Phys.Rev.Lett.87:(2001) 271301};
  A. De la Macorra    { JHEP01(2003)033};
  A. de la Macorra,{ Phys.Rev.D72:043508,2005}

\bib{IDE} L.~Amendola,
   Phys.\ Rev.\ D { 62}, 043511 (2000);
   M.~Kaplinghat and A.~Rajaraman, arXiv:astro-ph/0601517.
 D.~B.~Kaplan, A.~E.~Nelson and N.~Weiner,
  Phys.\ Rev.\ Lett.\  { 93}, 091801 (2004)
  R.~D.~Peccei,
  Phys.\ Rev.\ D { 71}, 023527 (2005)
 A.~W.~Brookfield, C.~van de Bruck, D.~F.~Mota and
  D.~Tocchini-Valentini, Phys.Rev.D { 73} (2006) 083515;
   M.~Kaplinghat and A.~Rajaraman, arXiv:astro-ph/0601517.
\bibitem{DAS}  S.~Das, P.~S.~Corasaniti and J.~Khoury,Phys.\ Rev.\ D { 73}, 083509 (2006)



\bib{IDE-n}   D.~B.~Kaplan, A.~E.~Nelson and N.~Weiner,
  Phys.\ Rev.\ Lett.\  { 93}, 091801 (2004)
  R.~D.~Peccei,
  Phys.\ Rev.\ D { 71}, 023527 (2005)
 A.~W.~Brookfield, C.~van de Bruck, D.~F.~Mota and
  D.~Tocchini-Valentini, Phys.Rev.D { 73} (2006) 083515;


\bib{IDE-ax}
  A.  de la Macorra, A. Melchiorri, P.  Serra, R.  Bean
astro-ph/0608351 (to be publsihed in Astrop.Phys.)

\bib{wapp}  S.~Das, P.~S.~Corasaniti and J.~Khoury,Phys.\ Rev.\ D { 73}, 083509 (2006)



\bib{ph.etc}
  S.~M.~Carroll, M.~Hoffman and M.~Trodden,
Phys.\ Rev.\ D { 68}, 023509 (2003); A. de la Macorra,  H. Vucetich,
{ JCAP09(2004)012}












\end{thebibliography}
\end{document}